%% file: main.tex
\documentclass[final]{svjour3}
\usepackage{graphicx}
\usepackage{rotating}
\usepackage{amssymb}
\usepackage{mathptmx}
\usepackage[numbers]{natbib}
\usepackage{wrapfig}
\makeatletter
\journalname{Journal of Low Temperature Physics}

\bibpunct{[}{]}{,}{n}{}{,}

\begin{document}

\def\todo#1{\textbf{#1}}

\newcommand{\hdblarrow}{H\makebox[0.9ex][l]{$\downdownarrows$}-}
\title{Development of MKIDs for measurement of the Cosmic Microwave Background with the South Pole Telescope}
\titlerunning{Development of MKIDs for measurement of the CMB with the South Pole Telescope}

\input{authors.tex}

\authorrunning{K. Dibert, et al. on behalf of the SPT Collaboration}

\maketitle

\begin{abstract}

We present details of the design, simulation, and initial test results of prototype detectors for the fourth-generation receiver of the South Pole Telescope (SPT). Optimized for the detection of key secondary anisotropies of the cosmic microwave background (CMB), SPT-3G+ will measure the temperature and polarization of the mm/sub-mm sky at 220, 285, and 345 GHz, beyond the peak of the CMB blackbody spectrum. The SPT-3G+ focal plane will be populated with microwave kinetic inductance detectors (MKIDs), allowing for significantly increased detector density with reduced cryogenic complexity. We present simulation-backed designs for single-color dual-polarization MKID pixels at each SPT-3G+ observation frequency. We further describe design choices made to promote resonator quality and uniformity, enabling us to maximize the available readout bandwidth. We also discuss aspects of the fabrication process that enable rapid production of these devices and present an initial dark characterization of a series of prototype devices.


\end{abstract}

\section{Introduction}

\begin{wrapfigure}{R}{0.4\textwidth}
\vskip -12pt
\includegraphics[trim=2.1in 0.0in 1.6in 0.0in,clip=true, width=0.5\textwidth]{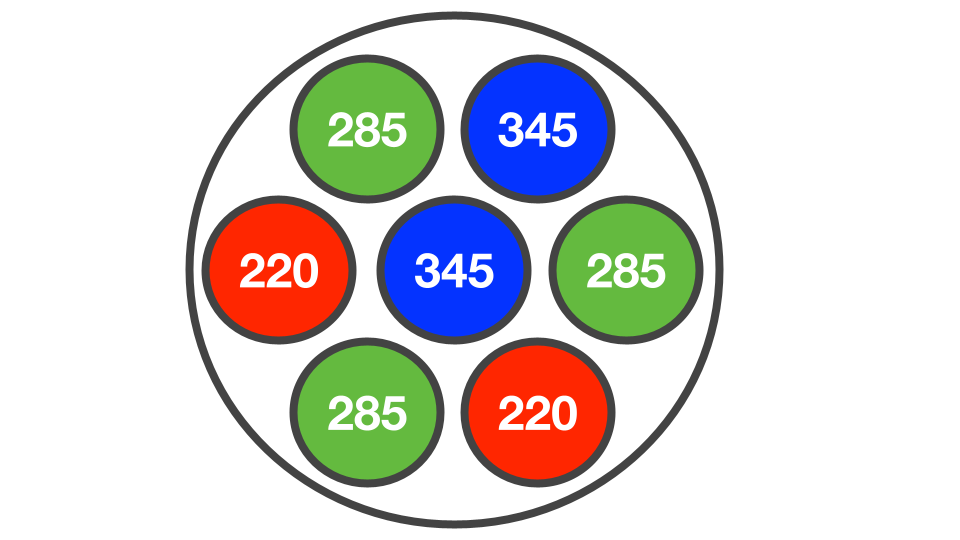}
\vskip -5pt
\caption{(color online) Schematic of the SPT-3G+ focal plane, which will consist of seven six-inch wafers, each containing nearly 5000 monochroic MKIDs observing at 220, 285, and 345 GHz.}
\vskip -12pt
\label{fig:plane}
\end{wrapfigure}

 The South Pole Telescope (SPT) is a 10-meter diameter mm/sub-mm wave telescope designed to measure the cosmic microwave background (CMB) and make new discoveries in high energy physics, astrophysics, and cosmology (e.g.,\cite{Balkenhol21, guns21, dutcher21}).  The SPT is currently equipped with the SPT-3G camera \cite{benson2014}, which is populated by an array of trichroic pixels made up of transition-edge sensor (TES) bolometers observing at 95, 150, and 220 GHz \cite{benson2014}. At the conclusion of the SPT-3G survey, the camera will be replaced by SPT-3G+, the fourth-generation receiver and focal plane, which is shown schematically in Figure \ref{fig:plane}. Observing the same 1500 square degree field as its predecessor, SPT-3G+ will enable high-sensitivity measurements of the temperature and polarization of the mm/sub-mm sky at observing frequencies of 225, 285, and 345 GHz, beyond the peak of the CMB blackbody spectrum. In combination with the SPT-3G data, SPT-3G+ will target new cosmological observables such as the patchy kinematic Sunyaev-Zel'dovich effect \cite{smith2017} and the Rayleigh scattering of the CMB \cite{lewis2013}. Robust detection of these observables will lead to groundbreaking constraints on the duration and timing of the epoch of reionization \cite{alvarez2021}, and provide new insights on the expansion of the universe just after recombination \cite{beringue2021}. While tantalizing from a science perspective, these signals are less than a few percent of the amplitude of the primary CMB signal, and their detection requires an experiment with sensitivity beyond that which is presently available.
 The sensitivity of superconducting TESs for CMB observation is no longer limited by individual detector noise, but by temporal fluctuations in the incident photon signal (photon noise).
Therefore, to achieve the increased sensitivity required to meet the SPT-3G+ science goals, a higher density of detectors on the focal plane is necessary. The detector density of the SPT-3G focal plane is now approaching a practical limit due to the number of wires that are needed to bias and readout each detector \cite{benson2014}. The drive for even greater detector density motivates the use of a more multiplexable CMB detector technology on the SPT-3G+ focal plane . 

Microwave kinetic inductance detectors (MKIDs) offer an attractive alternative solution, allowing for increased detector densities with significantly reduced readout complexity and more straightforward fabrication.
These advantages combine to enable the construction of large-format arrays of MKIDs with the necessary detector density for SPT-3G+ sensitivity requirements.
Due to their resonator-based design, MKIDs are highly multiplexable and have the further advantage that they require no additional cryogenic multiplexing electronics - each detector has a characteristic readout frequency set by the geometry of its resonator. The fundamental limit for the MKID multiplexing (MUX) factor is set by the quality factor $Q$ of the resonator, where a higher quality factor equates to a lower required readout bandwidth for a given detector. While individual MKIDs have been fabricated with extremely high quality factors \cite{vissers2010} \cite{auster2018}, the on-sky operation of a large MKID array presents important challenges with respect to the achievable channel densities. 
Previous mm-wave experiments have operated MKID arrays with detector counts of roughly 200 \cite{nika}, 1000 \cite{baselmans2017}, and 3000 \cite{galit2014}. The stage is now set for the creation of an operational mm-wave MKID array with tens of thousands of detectors.

SPT-3G+ will implement MKID arrays with a MUX factor approaching 2k channels per octave, which is a substantial increase over the current SPT-3G TES arrays.
As shown in Figure \ref{fig:plane}, the SPT-3G+ focal plane will be made up of seven single-color optics tubes each illuminating a single 150-mm MKID array. Each of the seven detector modules will contain close to 5k polarization-sensitive detectors, with the full focal plane containing 35k MKIDs in total. This will set new mm-wave MKID detector count and density milestones and will represent the next generation of MKID-based microwave sensing. The ultimate goal of this work will be to demonstrate full science-grade arrays of MKIDs at each SPT-3G+ frequency, which will then be integrated into the SPT-3G+ camera and eventually deployed to the South Pole for on-sky operation. We here present the designs, simulations, and initial test results of detector prototypes for the 220 GHz observing band.

\section{Pixel design}

\begin{figure}
    \includegraphics[trim=0.8in 0.0in 0.8in 0.0in,clip,width=0.40\textwidth]{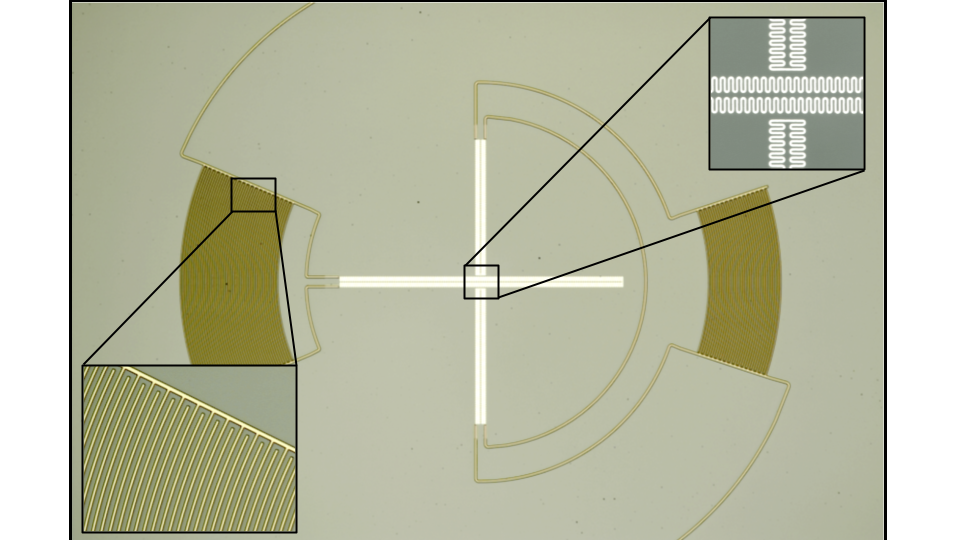}
    \includegraphics[width=0.59\textwidth]{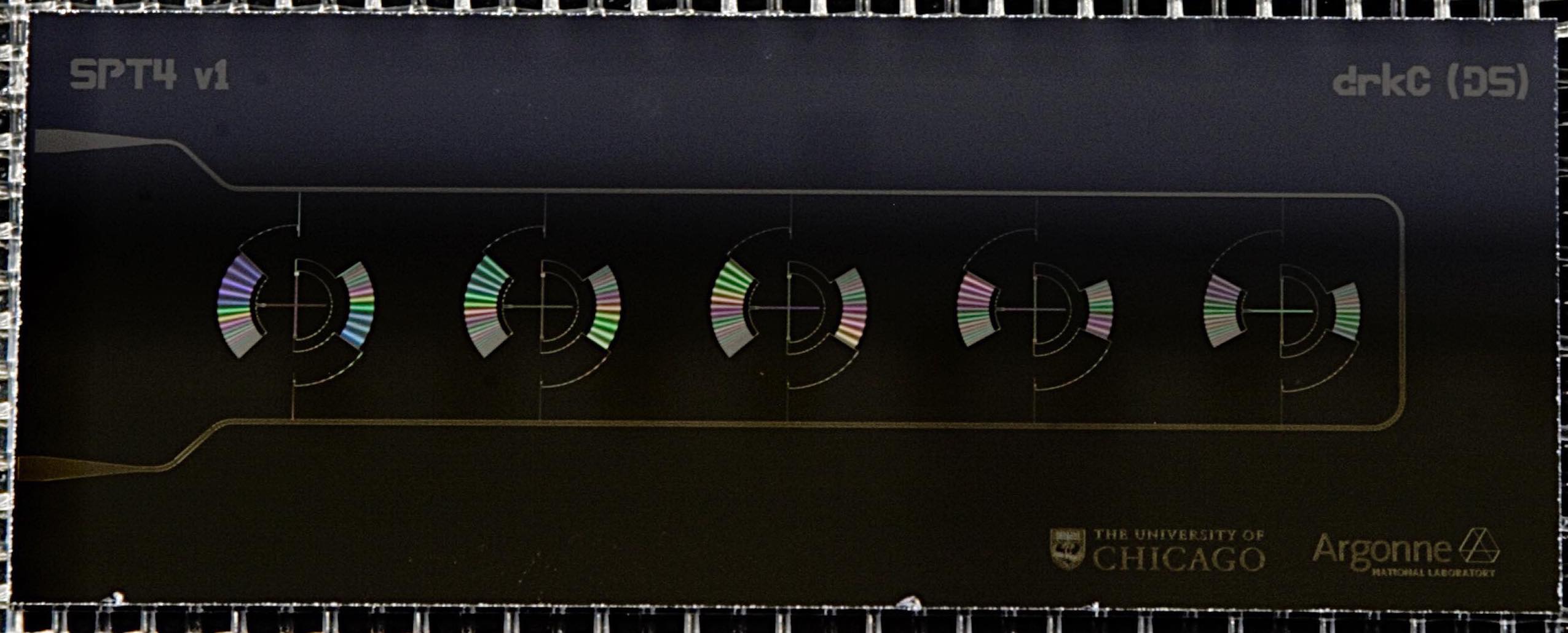}
    \caption{(color online) \textit{Left}: Microscope image of a single 220 GHz pixel, consisting of two MKIDs coupled to orthogonal polarization modes. Insets show detail on a niobium interdigitated capacitor and the aluminum inductors which double as optical absorbers. \textit{Right:} A prototype device consisting of five 220 GHz pixels.}
    \label{fig:photos}
\end{figure}

\begin{figure*}
    \centering
    \includegraphics[width=0.58\textwidth]{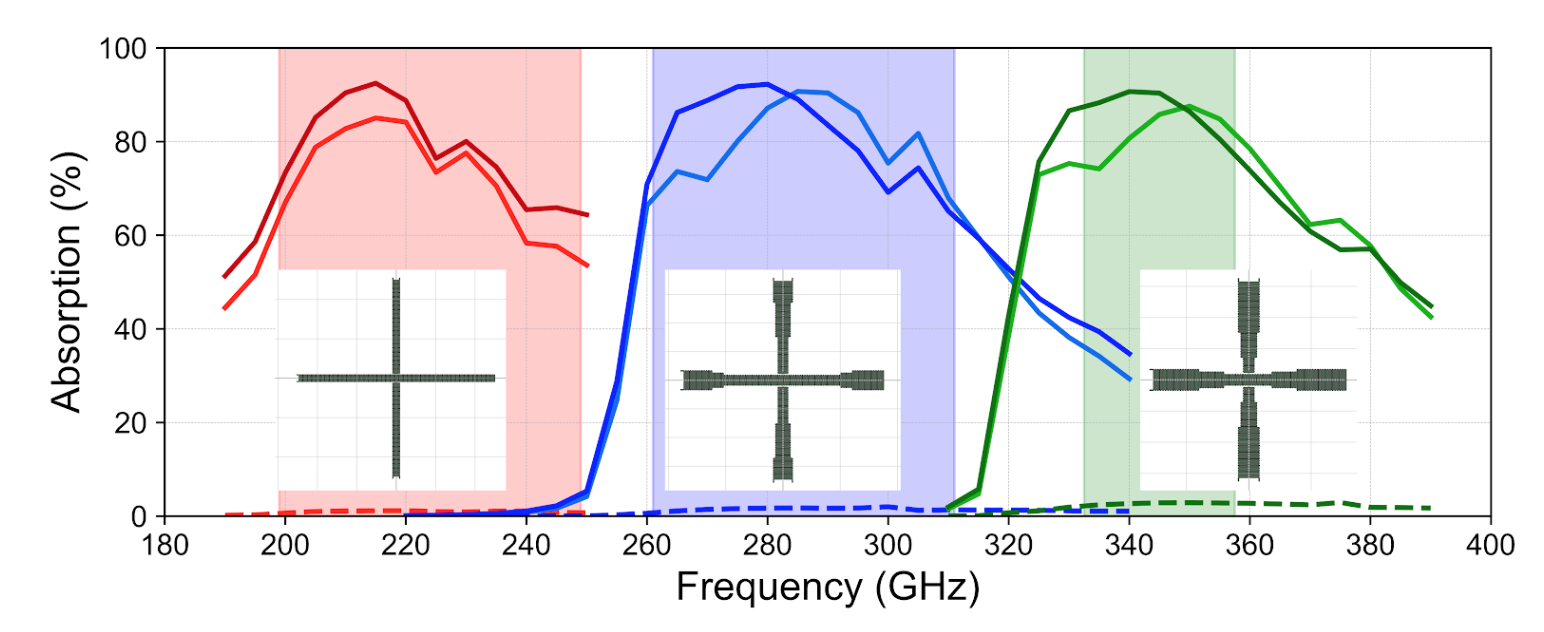}
    \includegraphics[width=0.41\textwidth]{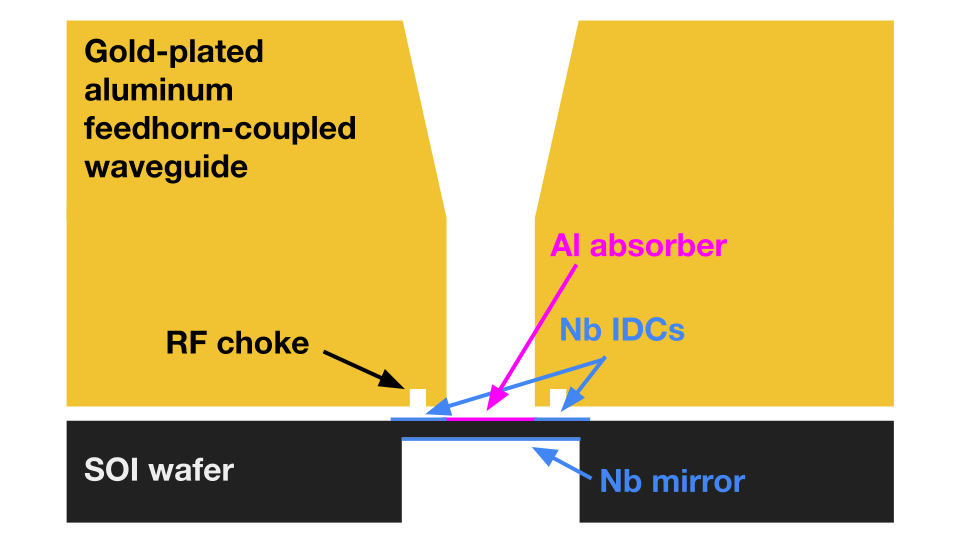}
    \caption{(color online) \textit{Left:} Simulated optical absorption for each SPT-3G+ frequency band and associated absorber geometry (inset). Shaded regions represent the SPT-3G+ bands, which will be defined by the waveguide cutoff at the lower edge and by a free-space metal-mesh filter at the upper edge. Dark and bright solid lines respectively indicate x and y-mode co-polarization absorption that is $\sim$80\% efficient in-band, while dashed lines indicate percent-level cross-polarization. \textit{Right:} Schematic depicting the absorber's position in relation to the aluminum waveguide, RF choke, and integrated silicon backshort.} 
\vskip -10pt
    \label{fig:hfss}
\end{figure*}

The pixel (shown in the left panel of Figure \ref{fig:photos}) is composed of two MKIDs aligned with orthogonal polarization modes. This design is based on previous work by \cite{vissers2020} and \cite{galit2014} with some important modifications that are necessary for operation at the South Pole. Each MKID consists of an aluminum inductor coupled to a niobium interdigitated capacitor (IDC). The inductor doubles as an impedance-matched optical absorber, where the geometry of the IDC is used to control the resonant frequency of the detector. To maximize the resonator quality factor $Q$ under the expected optical load, it is necessary to increase the total detector volume while retaining the required responsivity to stay photon-noise dominated. To achieve this, we have introduced meanders into the inductor (see inset of Figure \ref{fig:photos}) which has the effect of increasing both the volume and absorption efficiency. For higher observing frequencies (and hence smaller waveguides) the meanders increase in width towards the edges of the pixel where the electric field is weaker, allowing for the required increase in absorber volume while minimizing cross-polarization response. The resulting absorber shapes are shown in Figure \ref{fig:hfss}. With these optimized absorber designs, EM simulations indicate that high optical efficiency with low cross-polarization can be achieved (see simulated absorption curves in Figure \ref{fig:hfss}). Simulations indicate a loaded of $Q \sim 10^5$ for all frequency bands, allowing up to 5k channels in the 1-2 GHz readout bandwidth, well above our goal of 2k channels per octave. 
 
Each pixel is aligned to an aluminum circular waveguide which functions as a high-pass filter, and is encircled by an aluminum RF choke which constrains radiation leakage between adjacent pixels. To set the upper band edge, a metal-mesh low-pass filter is placed in front of a metallic smooth-walled feedhorn that guides radiation into the waveguide. Horn arrays are a well-established method of optical coupling to a MKID, with high efficiency demonstrated up to 1.2 THz \cite{groppi16}. In addition, metallic horn arrays form a natural shield that minimizes inter-pixel optical cross-talk that is known to be an issue for open MKID arrays \cite{yates2018}. 
The devices are patterned onto a silicon-on-insulator (SoI) wafer with a device layer thickness that is set such that the metallized backside of the wafer functions as an optical backshort. 
This integrated backshort has a number of advantages in terms of reducing mechanical complexity of the final detector arrays and assembly, as well as enabling post-fabrication adjustment of detector readout frequencies, as discussed in the following section.
The waveguide and silicon backshort setup is shown in the right panel of Figure \ref{fig:hfss}.

Figure \ref{fig:photos} shows single-pixel and full-chip photographs of a prototype 220 GHz device. The aluminum inductors and niobium IDCs are deposited and patterned via lithography on the front side of a SoI wafer with the necessary device layer thickness. A frontside silicon etch defines the alignment features and dicing lines, and a final backside silicon etch and niobium metallization completes the backshort. Requiring significantly fewer steps than the traditional multi-chroic architectures \cite{duff2016,posada2018,tang2020}, this process enables rapid design iteration between testing cooldowns. In order to minimize the number of RF readout channels and achieve our goal readout density of 2k/octave, accurate placement of the resonator frequencies is critical. The SPT-3G+ pixel design allows for post-fabrication adjustment of the resonator frequencies by performing a second lithography and trimming a small segment of the IDC. The use of an integrated silicon backshort makes this process possible and is key to maximizing detector yield. The technique has been successfully implemented by a number of groups \cite{Shu2018,McKenney2019}, and recently we have demonstrated the ability to place resonators with a fractional accuracy of $10^{-5}$ \cite{mcgeehan2018}. We have also successfully trimmed a prototype device with no loss in resonator yield.

\begin{figure}
\includegraphics[trim=0.25in 0.25in 0.0in 0.0in,clip=true, width=0.48\textwidth]{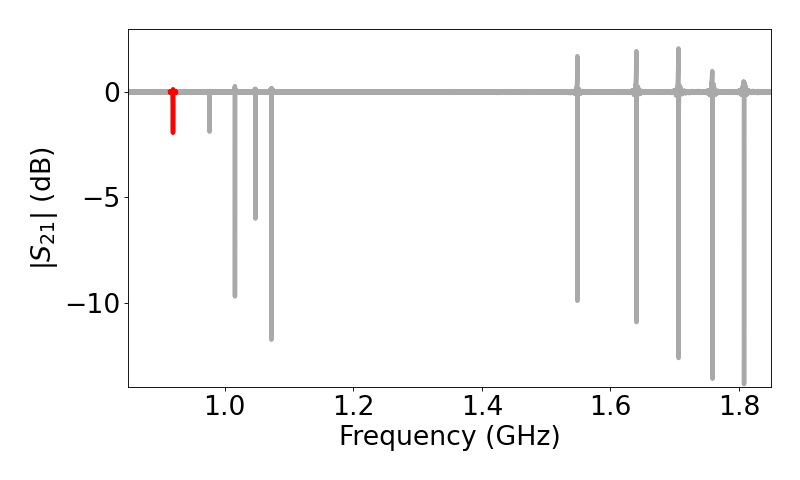}
\includegraphics[trim=0.25in 0.25in 0.0in 0.0in,clip=true, width=0.48\textwidth]{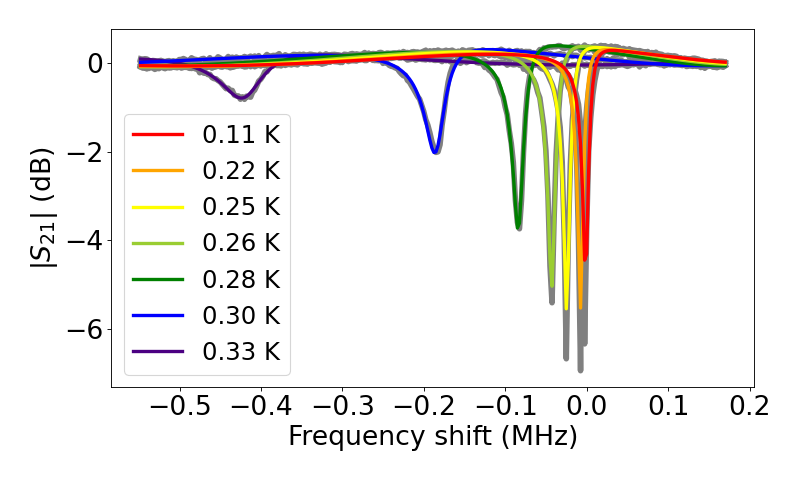}
\caption{(color online) \textit{Left:} Dark frequency sweep of a 10-pixel chip identical to that shown in the right-hand panel of Figure \ref{fig:photos}. Resonances are arranged into two banks across the 1 GHz readout range, with the lower-frequency bank containing x-polarization-sensitive resonators, and the higher-frequency bank containing y-polarization-sensitive resonators. A high-pass filter has been applied to the VNA baseline for aesthetic purposes.  \textit{Right:} Temperature response of the resonance indicated in red on the left-hand panel. Data is shown in gray, and colored lines represent Lorentzian fits to the data at each temperature. These fits indicate inductor quality factors ranging from $\sim 2 \times 10^{5}$ at 110 mK to $\sim 2 \times 10^{4}$ at 330 mK.}
\vskip -14pt
\label{fig:dark}
\end{figure}

\section{Initial test results}

The result of a dark characterization of a chip identical to that pictured in the right-hand panel of Figure \ref{fig:photos} is shown in the left-hand panel of Figure \ref{fig:dark}. The ten resonators present in this chip design appear within the expected readout range, and are located within two banks as intended. The first bank of five contains x-polarization-sensitive resonators, while the second bank contains y-polarization-sensitive resonators. The right-hand panel of Figure \ref{fig:dark} shows the response to temperature of the lowest-frequency resonance on this chip (indicated in red in the left-hand panel of Figure \ref{fig:dark}). As expected, the resonant frequency shifts downward with increasing temperature, and the resonance widens. Lorentzian fits to the data at each temperature (colored lines) indicate inductor quality factors ranging from $\sim 2 \times 10^5$ at 110 mK to $\sim 2 \times 10^{4}$ at 330 mK. Assuming a 110 mK operating temperature, this agrees with simulations and is well within our target quality factor range, indicating that our ideal 2k/octave multiplexing goal is feasible. 

\begin{figure}
\centering
\includegraphics[trim=0.0in 0.0in 0.0in 1.2in,clip=true, width=0.8\textwidth]{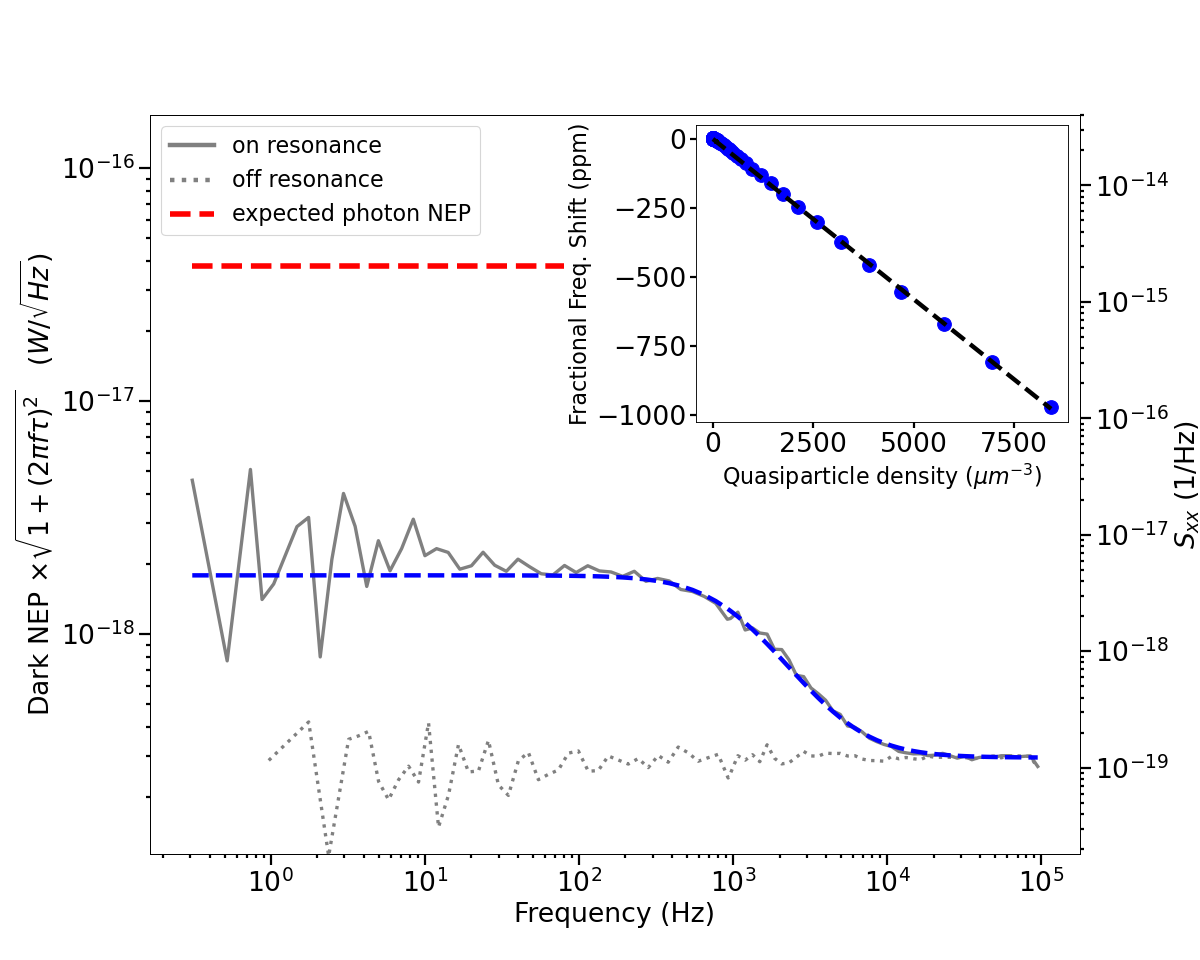}
\vskip -10pt
\caption{(color online) The left axis shows the dark NEP corrected for the responsivity roll off (grey) for an aluminum-only device as compared to the expected photon NEP (red dashed) with 5 pW optical loading, indicating background domination in this optimized ($\eta_{opt} = 1$) scenario. The dark NEP was calculated from the dark responsivity (see inset) and the detector noise. On the right axis, the grey lines are power spectral densities of on and off-resonance noise timestreams measured at 110mK. The blue dashed line is a fit to the on-resonance dark detector noise, which is limited by the generation-recombination noise of the device. \textit{Inset:} Linear fit to fractional frequency shift vs quasiparticle number density. The slope of this fit was used to calculate the dark responsivity of the resonator.}
\vskip -10pt
\label{fig:dark_response}
\end{figure}

Figure \ref{fig:dark_response} shows the power spectral density ($S_{xx}$) of a series of time ordered data measured for the same resonator at a temperature of 110 mK. We observe the form expected for an MKID that is limited by the generation-recombination noise associated with the superconducting charge carriers \cite{deVisser2012}. The left axis shows the conversion of this power spectral density to a dark NEP. For a lumped element MKID architecture, this is known to be equivalent to the NEP referenced to absorbed optical power \cite{Janssen2014}. The dark NEP is estimated from \cite{baselmans2008} as
\begin{equation}
    NEP = \sqrt{S_{xx}} \Big[ (\frac{1}{f_0} \frac{\delta f_0}{\delta n_{qp}} ) \frac{\delta n_{qp}}{\delta P_{opt}} \Big]^{-1},
\end{equation}
where the product in the brackets represents dark responsivity. The term in parentheses is obtained by performing dark sweeps of the resonance at multiple temperatures. The expected quasiparticle density is calculated from each temperature via:
\begin{equation}
    n_{qp} = 2N_0\sqrt{2\pi k_B T \Delta} e^{-\frac{\Delta}{k_BT}},
\end{equation}
where $N_0 = 1.73 \times 10^{10} \, \mu \textrm{m}^{-3} \textrm{eV}^{-1}$ is the single-spin density of states for aluminum, $k_B$ is the Boltzmann constant, $T$ is the temperature of the resonator, and $\Delta = 1.76 k_B T_c$, with $T_c = 1.4$ K for the aluminum inductor. The critical temperature $T_c$ was extracted from a fit to the resonant frequency shift versus device temperature. Each blue point on the inset to Figure \ref{fig:dark_response} represents the fractional frequency shift of the device in the presence of a given quasiparticle density. The value of $\frac{1}{f_0}\frac{\delta f_0}{\delta n_{qp}}$ is then the magnitude of the slope of the fit to $\frac{\delta f_0}{f_0}$ vs $n_{qp}$. This fit is shown in the inset to Figure \ref{fig:dark_response}, and was performed over a high temperature range where the quasiparticle response is dominant. The remaining term in the brackets, from \cite{baselmans2008}, is
\begin{equation}
    \frac{\delta n_{qp}}{\delta P_{opt}} = \frac{\eta_e \tau_{qp}}{\Delta V_L},
\end{equation}
where $\eta_e \sim 0.8$ is the expected pair breaking efficiency, $V_L$ is the inductor volume, and $\tau_{qp} = 220 \, \, \mu$s is the quasiparticle lifetime, which is derived from the fit to the dark noise power spectral density rolloff (blue dashed line in Figure \ref{fig:dark_response}) at 110 mK. The fit to the noise rolloff here is well represented by a single-pole Lorentzian with time constant $\tau_{qp}$, as expected since the the measured quasiparticle lifetime is significantly longer than the resonator ringdown time $\tau_{res} = 2Q_r/\omega_0 \approx 20 \, \, \mu$s.

The red dashed line in Figure \ref{fig:dark_response} represents the expected NEP due to photon noise with 5 pW optical loading, which is the estimated loading at 220 GHz due to atmospheric and instrument emissivity. Within the resonator bandwidth, this photon noise limited NEP is given by
\begin{equation}
    NEP^2_{ph} = 2\eta_{opt}P_{opt}h\nu(1+\eta_{opt}\bar{n}_{ph}),
\end{equation}
where $\bar{n}_{ph} = [e^{h\nu/k_BT} -1]^{-1}$ is the mean photon occupation number and $\eta_{opt}=1$ is the best-case detector optical efficiency. This estimate suggests that the intrinsic device noise is well below the expected photon NEP at 220 GHz, and provides strong evidence that this detector design is well suited to achieve the sensitivity requirements of SPT-3G+. 

\section{Conclusions}

We have presented the design for a polarization-sensitive, feedhorn-coupled pixel for use on SPT-3G+, the next-generation focal plane of the South Pole Telescope. We have fabricated prototypes of the 220 GHz design, and have tested these prototypes in a dark environment. Resonator placement and quality factors align with simulations, indicating the feasibility of our 2k/octave multiplexing goal. The dark NEP, calculated assuming a best-case-scenario of 100\% optical efficiency, falls well below the theoretical photon NEP at 5 pW of loading, suggesting that background domination is achievable with an optical efficiency above $\sim20\%$. Optical testing will provide further information on the optical efficiency of these devices, and is currently
underway. Further work remains to produce full arrays at all three SPT-3G+ frequencies and to integrate these with the SPT-3G+ cryostat for on-sky deployment.

\begin{acknowledgements}
This material is based upon work supported by the National Science Foundation under NSF-1852617 and NSF-2117894. Work at Argonne National Laboratory was supported by the U.S. Department of Energy (DOE), Office of Science, Office of High Energy Physics, under contract DE-AC02-06CH1137. Partial support was also provided by the DOE Graduate Instrumentation Research Award. This work made use of the Pritzker Nanofabrication Facility of the Institute for Molecular Engineering at the University of Chicago, which receives support from Soft and Hybrid Nanotechnology Experimental (SHyNE) Resource (NSF ECCS-2025633), a node of the National Science Foundation’s National Nanotechnology Coordinated Infrastructure.
\end{acknowledgements}

\pagebreak

\bibliographystyle{plain}
\bibliography{main.bib}

\vskip 20pt
\noindent \textbf{Data availability statement}: The datasets generated during and/or analyzed during the current study are available from the corresponding author on reasonable request.

\end{document}

%% file: authors.tex

\author{
K.~Dibert\textsuperscript{1},
P.~Barry\textsuperscript{2},
Z.~Pan\textsuperscript{2},
A.~Anderson\textsuperscript{3,4},
B.~Benson\textsuperscript{1,3,4}, 
C.~Chang\textsuperscript{1,2,3}, 
K.~Karkare\textsuperscript{3,4},
J.~Li\textsuperscript{2},
T.~Natoli\textsuperscript{3}, 
M.~Rouble\textsuperscript{5},
E.~Shirokoff\textsuperscript{1,3}, 
and~A.~Stark\textsuperscript{6},
on behalf of the South Pole Telescope Collaboration
}


\institute{
\textsuperscript{1}{Dept. of Astronomy \& Astrophysics, U. Chicago, 5640 South Ellis Avenue, Chicago, IL, 60637, USA} \\
\textsuperscript{2}{Argonne National Laboratory, 9700 South Cass Avenue., Argonne, IL, 60439, USA} \\
\textsuperscript{3}{Kavli Institute for Cosmological Physics, U. Chicago, 5640 South Ellis Avenue, Chicago, IL, 60637, USA} \\
\textsuperscript{4}{Fermi National Accelerator Laboratory, MS209, P.O. Box 500, Batavia, IL, 60510, USA} \\
\textsuperscript{5}{Department of Physics, McGill University, 3600 Rue University, Montreal, Quebec H3A 2T8, Canada} \\
\textsuperscript{6}{Harvard-Smithsonian Center for Astrophysics, 60 Garden Street, Cambridge, MA, 02138, USA} \\
}